**Nanoscale-confined Terahertz Polaritons in a van der Waals Crystal**

*Thales V. A. G. de Oliveira,\* Tobias Nörenberg, Gonzalo Álvarez-Pérez, Lukas Wehmeier, Javier Taboada-Gutiérrez, Maximilian Obst, Franz Hempel, Eduardo J. H. Lee, John M. Klopf, Ion Errea, Alexey Y. Nikitin, Susanne C. Kehr,\* Pablo Alonso-Gonzaléz,\* and Lukas M. Eng.*

Dr. T. V. A. G. de Oliveira, T. Nörenberg, L. Wehmeier, M. Obst, F. Hempel, Dr. S. C. Kehr, and Prof. L. M. Eng
Institut für Angewandte Physik, Technische Universität Dresden
Dresden 01187, Germany
E-mail: thales.oliveira@tu-dresden.de, susanne.kehr@tu-dresden.de

Dr. T. V. A. G. de Oliveira, T. Nörenberg and Prof. L. M. Eng
Dresden-Würzburg Cluster of Excellence-EXC 2147 (ct.qmat)
Dresden 01062, Germany

G. Álvarez-Pérez, J. Taboada-Gutiérrez and Dr. P. Alonso-González
Department of Physics
University of Oviedo
Oviedo 33006, Spain
E-mail: pabloalonso@uniovi.es

G. Álvarez-Pérez, J. Taboada-Gutiérrez and Dr. P. Alonso-González
Center of Research on Nanomaterials and Nanotechnology
CINN (CSIC−Universidad de Oviedo)
El Entrego 33940, Spain

Dr. E. J. H. Lee
Departamento de Física de la Materia Condensada
Condensed Matter Physics Center (IFIMAC)
Universidad Autónoma de Madrid
Madrid 28049, Spain

Dr. T. V. A. G. de Oliveira and J. M. Klopf
Institute of Radiation Physics
Helmholtz-Zentrum Dresden-Rossendorf
Dresden 01328, Germany

Dr. I. Errea
Fisika Aplikatua 1 Saila
University of the Basque Country (UPV/EHU)
Donostia/San Sebastián 20018, Spain

Dr. I. Errea
Centro de Física de Materiales (CSIC-UPV/EHU)
Donostia/San Sebastián 20018, Spain

Dr. I. Errea and Dr. A. Y. Nikitin
Donostia International Physics Center (DIPC)
Donostia/San Sebastián 20018, Spain

Electromagnetic field confinement is crucial for nanophotonic technologies, since it allows for enhancing light-matter interactions, thus enabling light manipulation in deep sub-wavelength scales. In the terahertz (THz) spectral range, radiation confinement is conventionally achieved with specially designed metallic structures – such as antennas or nanoslits – with large footprints due to the rather long wavelengths of THz radiation. In this context, phonon polaritons – light coupled to lattice vibrations – in van der Waals (vdW) crystals have emerged as a promising solution for controlling light beyond the diffraction limit, as they feature extreme field confinements and low optical losses. However, experimental demonstration of nanoscale-confined phonon polaritons at THz frequencies has so far remained elusive. Here, we provide it by employing scattering-type scanning near-field optical microscopy (s-SNOM) combined with a free-electron laser (FEL) to reveal a range of low-loss polaritonic excitations at frequencies from 8 to 12 THz in the vdW semiconductor α-MoO$_3$. We visualize THz polaritons with i) in-plane hyperbolic dispersion, ii) extreme nanoscale field confinement (below $\lambda_o/75$) and iii) long polariton lifetimes, with a lower limit of > 2 ps.

Phonon polaritons (PhP)[1–3] in polar dielectrics feature very low optical losses due to the reduced rate of electronic scattering,[4] and their response can be readily adjusted by size scaling (e.g. number of layers in a vdW crystal)[5] or ion intercalation.[6,7] They are particularly interesting in so-called hyperbolic media (those whose dielectric permittivities have opposite signs along different crystallographic directions), where they exhibit a strongly directional behavior. The polaritonic iso-frequency curves (IFCs, slices of the dispersion surface in frequency-momentum space by a plane of a constant frequency $\nu$) are described by open hyperbolas, giving rise to exotic and very intriguing optical phenomena, such as extremely high momenta[8] (as needed for electromagnetic field confinement), small group velocities, negative phase velocities[9] (with great potential for applications exploiting negative refraction and Doppler effects at the nanoscale[10,11]), ultra-long lifetimes[3,12] and, more recently, flat-band canalization of topological polaritons in twisted vdW bilayers.[13–16]

Nevertheless, hyperbolic PhPs only exist within spectral intervals that are defined by the material itself: the so-called *reststrahlen bands* (RB). RBs in polar dielectrics, located within the transverse-optical (TO) and longitudinal-optical (LO) phonon frequencies, are narrow in spectral width and, moreover, naturally accompanied by losses. To date, observation of low-loss hyperbolic PhPs remains restricted to a few mid-infrared bands.[6,12,17,18] Therefore, finding low-loss polaritonic bands collectively spanning the full electromagnetic spectrum, as well as strategies to tailor their spectral position[6] is urgently needed. Particularly in the long-wavelength regime, novel strategies to enhance and confine THz fields to nanoscale dimensions are also highly desired, complementing or even replacing large footprint metallic antennas. Applications could be envisioned for example to enhance the nonlinear THz frequency conversion efficiency in Dirac materials[19] or to provide a down-scalable route for the generation of intense THz transients in spin-switching devices.[20]

Here, we demonstrate the existence of nanoscale-confined, low-loss phonon polaritons at THz frequencies in the biaxial vdW crystal α-MoO$_3$. To do this, we combine s-SNOM nanoimaging with a tunable infrared FEL (sketched in **Figure 1**a), making use of its sub-millielectronvolt energy resolution (Figure 1b and Methods Section) to access the fine structure of the polariton dispersion. We experimentally demonstrate the existence of THz polariton bands in α-MoO$_3$, and corroborate our findings by several theoretical approaches. Our real-space visualizations reveal two THz PhP bands with in-plane hyperbolic anisotropy, orthogonal propagation directions, exceptional confinement factors, and low losses – exhibiting lifetimes of 3.1 ± 0.4 ps and 9 ± 4 ps for polaritons propagating along the [001] and [100] crystallographic axes, respectively.

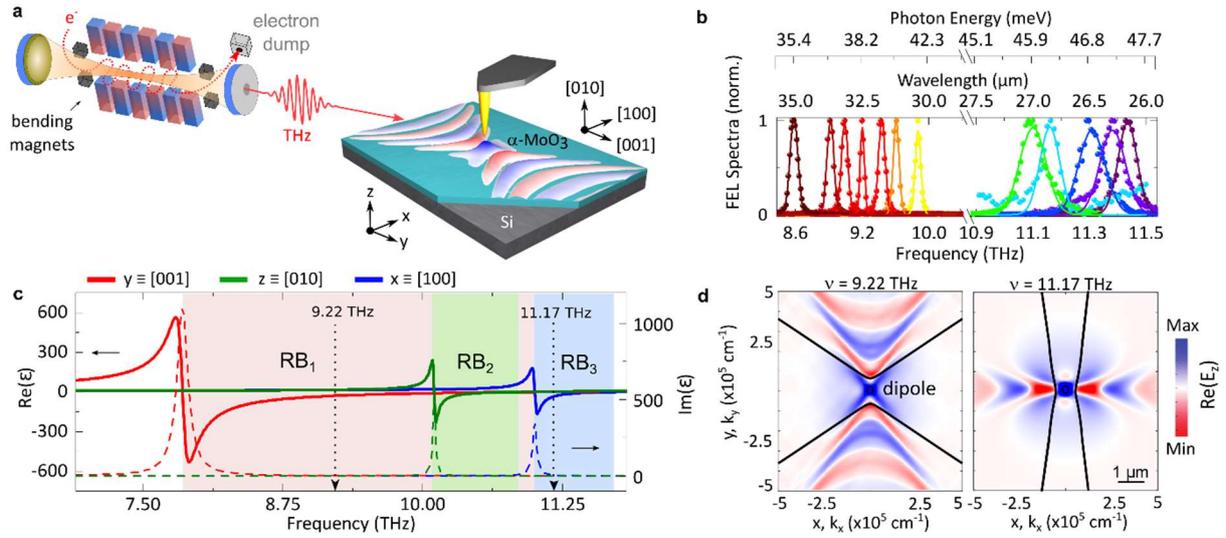

**Figure 1. Polariton nanoimaging with a free-electron laser coupled to a s-SNOM microscope and prediction of THz polaritons in α-MoO₃ with hyperbolic propagation.** (a) Illustration of the experiment. The s-SNOM tip is polarized by intense picosecond THz pulses provided by a widely tunable free-electron laser, which in turn launches polaritons in an α-MoO₃ slab that propagate away from the tip. (b) Spectra of the FEL-pulses employed in the experiment for the reststrahlen bands RB₁ and RB₃ of α-MoO₃ as indicated in (c). The data points (dots) were obtained by grating spectrometry and fitted using Gaussian distributions (lines). (c) Dielectric permittivity tensor of α-MoO₃ in the THz spectral range obtained by correlating ab-initio calculations with near- and far-field experiments (Supplementary Note S6). The real (solid lines) and imaginary (dashed lines) parts of the permittivity tensor reveal three distinct reststrahlen bands with negative permittivities along different crystal axes, shaded in red ([001]), green ([010]), and blue ([100]). (d) Open hyperbolic polaritonic IFCs in momentum space $(k_x, k_y)$ for the frequencies marked in (c) (black lines), overlaid with the numerical simulations of the electric field distribution $Re(E_z)$ in real-space $(x, y)$ (false color plots).

The highly-asymmetric, biaxial crystal structure of α-MoO₃ (see Supplementary Note S1 for sample details) gives rise to different dielectric permittivities along all three crystallographic directions, in addition to strongly anisotropic Raman vibrations over the whole infrared (IR) spectrum (see Supplementary Note S3 for polarization-resolved Raman spectroscopy characterization). As a polar crystal, α-MoO₃ has its LO-TO phonon degeneracy lifted for a variety of IR-active vibration modes, defining bands of high reflectivity (the RBs), wherein the real part of the frequency-dependent permittivity tensor becomes negative $Re(\varepsilon_i) < 0$ ($i = x, y, z$). Polaritons emerging at high-frequencies (ν > 12 THz) in α-MoO₃ have been observed within a few RBs[12,18,21,22] resulting from interatomic stretching vibrational modes.

Interestingly, several bending vibrational modes also exist[23,24] at the lower-frequency side of the spectrum ($\nu$ < 12 THz). To explore the THz response dictated by these modes, we obtained the dielectric permittivity tensor of α-MoO$_3$ (so far unknown in the THz spectrum) by correlating ab-initio calculations and near-field polaritonic experiments, as described in Supplementary Notes S5 and S6. As shown in Figure 1c, we observe spectral bands (marked as RB$_{1-3}$) wherein at least one of the principal components of the permittivity is negative. In particular, within the bands RB$_1$ and RB$_3$ this occurs along the [001] and [100] crystal directions, respectively, indicating their potential to support PhPs with in-plane hyperbolic propagation. Note that for α-MoO$_3$ the vdW layers are stacked along the [010] direction, which defines the z-coordinate as indicated in Figure 1a.

To further study this polaritonic response, we performed analytical and numerical analyses based on electromagnetic theory (details in Methods Section) at targeted excitation frequencies for our sample system, namely thin-slabs of α-MoO$_3$ placed on a high-resistivity silicon substrate. The substrate shows ultra-low losses at frequencies < 12 THz (permittivity data in Supplementary Note S4), and was chosen in order to minimize the impact of the dielectric environment[25–27] on the polaritonic response. The results are shown in Figure 1d (black continuous lines), where we predict the existence of highly anisotropic polaritons by calculating their IFCs for a thin α-MoO$_3$ slab with representative thickness $d$ = 197 nm. We directly obtain hyperbolic IFCs exhibiting accessible wavevectors within hyperbolic sectors centered along the crystallographic directions [001] (for RB$_1$ excited with $\nu$ = 9.22 THz) and [100] (for RB$_3$ excited with $\nu$ = 11.17 THz), corroborating the existence of in-plane hyperbolic polaritons in these spectral bands. To better visualize the propagation and orthogonality of these polaritons, we also performed full-wave electromagnetic simulations and extracted the spatial distribution of the vertical component of the electric field Re[$E_z(x, y)$] (overlaid as color plots over Figure 1d panels). We observe the characteristic features of hyperbolic PhPs, such as concave wave-fronts, ray-like directional propagation, and significantly reduced wavelengths (as compared to the wavelength in free space). Specifically, evaluation of these polaritonic wavelengths yield values of 2.149 ± 0.005 μm and 2.705 ± 0.005 μm in RB$_1$ and RB$_3$, respectively (referred to as PhP$_{[001]}$ and PhP$_{[100]}$ hereafter).

We verify our theoretical predictions by performing s-SNOM nanoimaging experiments of an α-MoO$_3$ flake (**Figure 2**a) with thickness $d$ = 197 nm at the selected THz frequencies highlighted in Fig. 1c. To this end, we employ polariton interferometry nanoimaging[17,28] for the first time using FEL pulses. The raster-scanned s-SNOM tip acts as an antenna providing near-fields with the necessary momenta to launch polariton pulses that propagate away from the tip, reflect on sharp flake edges and return to the tip, where they are re-scattered into the far-field for detection. The measured signal is then modulated by the self-interference of the forward- (tail) and backward (head) propagating polariton pulse, thus allowing us to directly access in real-space its wavelength and decay lengths. Figures 2b,c show near-field intensity ($S_{2\Omega}$) images for the flake shown in Figure 2a at illumination frequencies $v$ = 9.22 THz and $v$ = 11.17 THz, targeting the predicted polaritonic responses of Figure 1d. Our near-field images show periodic $S_{2\Omega}$ signals (fringes) parallel to specific flake edges. Particularly, at $v$ = 9.22 THz (Figure 2b and its associated profiles at the top panel of Figure 2d), such fringes appear exclusively at the bottom edge, which is oriented orthogonally to the [001] crystal axis. This result directly reveals PhPs propagating with strongly in-plane anisotropic character. Notably, at this particular frequency we experimentally obtain a polariton wavelength of $\lambda_{\text{PhP}_{[001]}}^{v\ =\ 9.22\ \text{THz}}$ = 2.82 ± 0.08 μm, which is substantially smaller than the free-space wavelength $\lambda_o$ = 32.5 μm, providing clear evidence of the deep-subwavelength character of these THz polaritons. At $v$ = 11.17 THz (Figure 2c and its associated profiles at the bottom panel of Figure 2d), a similar anisotropic response is observed, with fringes appearing along one specific edge of the flake, but in this case orthogonal to the [100] crystal axis, again revealing the excitation of highly directional, in-plane PhPs at THz frequencies. Particularly, at this frequency we observe a polariton wavelength of $\lambda_{PhP_{[100]}}^{\nu=11.17\,THz} = 3.9 \pm 0.9$ μm, which is again much smaller than the free-space wavelength $\lambda_o = 26.8$ μm. These experimentally extracted polariton wavelengths are in good agreement with the numerically predicted quantities discussed above.

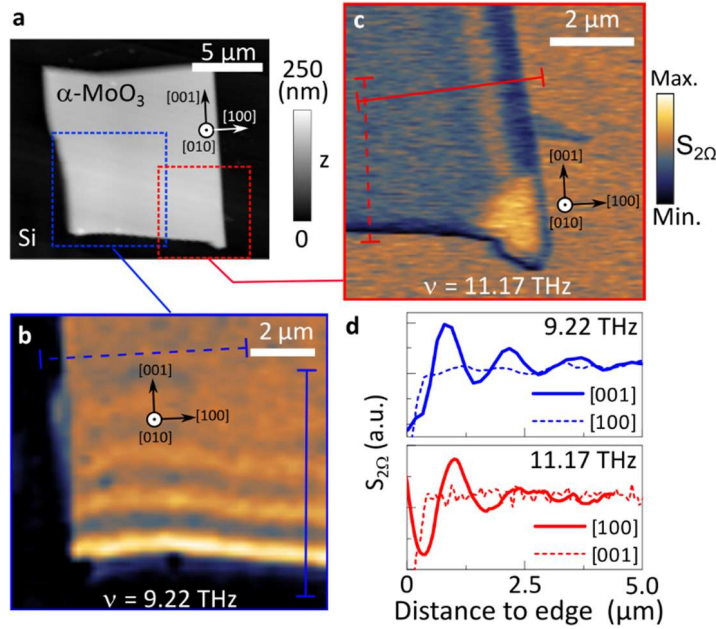

**Figure 2. Near-field visualization of THz polaritons**. (a) Atomic force microscopy image of an α-MoO₃ flake studied in this work with thickness d = 197 nm. The dashed boxes denote the areas where the near-field images displayed in (b) and (c) have been extracted. (b),(c) Representative near-field intensity $S_{2\Omega}$ images taken at THz frequencies within RB₁ (ν = 9.22 THz) and RB₃ (ν = 11.17 THz), respectively. (d) Polariton profiles extracted from the color-coded positions marked in (b,c). THz polaritons appear exclusively at the edges orthogonal to their principal propagation direction.

To analyze in detail the properties and tunability of THz polaritons in α-MoO₃, we quantitatively extract their dispersion on flakes with different thicknesses using a wide set of FEL frequencies (Figure 1b). In **Figure 3**a, we plot the experimental and analytical dispersion relations $\nu(\mathbf{k})$ for PhPs propagating along the [001] crystal direction (PhPs[001], symbols and analytical curve shaded in red) and along the [100] crystal direction (PhP[100], symbols and analytical curve shaded in blue). These measurements were performed on a set of four flakes with thicknesses $d$ = 53 nm, 131 nm, 197 nm, and 295 nm. The analytical dispersions derived from Equation S8 (Supplementary Note S6) show excellent agreement with the experimental data. Since we observe increasing ***k*** values (smaller wavelengths) by sweeping the illumination frequency from $\nu_{TO} \rightarrow \nu_{LO}$ (increasing frequency), we can conclude that the nanoimaged THz polaritons in this frequency range propagate with a positive phase velocity, in analogy with the mid-IR polaritons reported previously.[12] We stress that no discernible polariton fringes are observed at edges parallel to the main propagation direction of PhPs[001] and PhPs[001] crystal axes, thus providing strong evidence for their in-plane anisotropic character. Such anisotropic response can be visualized by inspecting the atomic displacement vectors associated with the vibrational modes of our polaritons, as obtained by ab-initio calculations (Figure 3b, for further details on the mode assignments see Supplementary Note S5). For PhPs[001], we observe a bending deformation mode involving three light oxygen atoms and the heavy molybdenum cation ($Mo^{6+}$), wagging along the [001] axis. For PhPs[100], a simpler vibrational mode is observed, wherein only the tri-coordinated oxygen atoms wag along the [100] axis. Note that for the latter, an out-of-plane vibration mode is also observed along the [010] crystal direction, though its resulting net dipole moment is zero (i.e. Raman active), thus not contributing directly to the anisotropic polaritonic response.

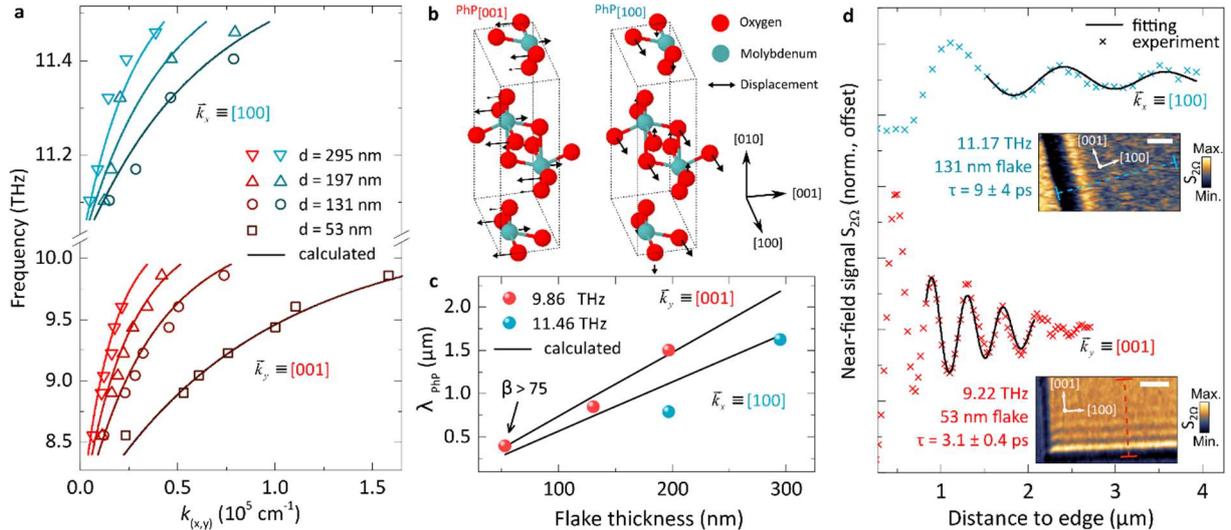

**Figure 3. Dispersion, lifetime and tunability of THz polaritons in α-MoO₃.** (a) THz polariton dispersion of α-MoO₃ flakes of different thicknesses d. Data points and analytical dispersion are color-coded in shades of blue (for PhP[100]), and red (for PhP[001]). (b) Illustration of the ab-initio atomic displacements of α-MoO₃ at the TO frequency associated with the RB₁ ($\nu_{adjusted}$ = 7.85 THz), and with the RB₃ ($\nu_{adjusted}$ = 11.00 THz) (c) Experimental (spheres) and analytical (solid line) polariton wavelengths as function of sample thickness. The highest

confinement factor β > 75 is achieved for the thinnest flake at THz frequencies close to the LO phonon modes. (d) Representative line-profiles for PhP$_{[100]}$ (blue curve, $d$ = 131 nm, ν = 11.17 THz) and PhP$_{[001]}$ (red curve, $d$ = 53 nm, ν = 9.22 THz) extracted from the near-field $S_{2\Omega}$ images in the inset. The curves are normalized to the average $S_{2\Omega}$ signal over the α-MoO$_3$ sample and offset for sake of clarity. Inset scale bars: 1 µm.

The wide tunability of these THz polaritons has been already indicated in the thickness-dependent dispersion relations of Figure 3a. In order to evaluate in a more explicit way the highest polariton confinement factor obtained (β = λ$_o$/λ$_{PhP}$), we plot the polariton wavelength as a function of the flake thicknesses for fixed illumination frequencies (9.86 THz for PhP$_{[001]}$ and 11.46 THz for PhP$_{[100]}$), as shown in Figure 3c. A linear scaling law is observed for these polaritons, in good agreement with our analytical calculations (black lines). Experimentally, we obtain wavelengths reaching the nanoscale for the thinnest flake, namely down to $\lambda_{PhP_{[001]}}^{\nu\,=\,9.86\,THz} = 397 \pm 13$ nm when illuminating with $\lambda_o = 30.4$ µm, thus exhibiting a confinement factor $\beta > 75$. Such extraordinary spatial confinement indicates that the field strength is strongly enhanced in the vicinity of such a polaritonic element, as dictated by energy-flux conservation.

Additionally, remarkably shallow dispersion slopes are observed due to the high $k$ values and relatively narrow TO-LO frequency splitting. Consequently, extraordinary slow polaritonic group velocities are obtained (v$_{g,i}$ = ∂ω/∂k$_i$, with ω = 2πν being the illumination angular frequency. See Supplementary Note S2 for details). For the flake with $d$ = 131 nm (circle data points in Fig 3a), we obtain $v_g$ of around $2.2 \times 10^{-3}c$ (PhP$_{[001]}$ at ν = 9.86 THz) and $0.5 \times 10^{-3}c$ (PhP$_{[100]}$ at ν = 11.4 THz) for polaritons with comparable momenta. We note that polariton visualization becomes increasingly difficult the slower it propagates, as this entails that for a certain illumination bandwidth a large range of $\boldsymbol{k}$ vectors are simultaneously excited, therefore leading to self-interference effects and limited apparent propagation.[29] For the PhP$_{[100]}$ modes manifested in the ultra-narrow RB$_3$ (ν$_{LO}$ − ν$_{TO}$ ≈ 0.72 THz), even with our relatively narrowband FEL excitation (Fig. 1b), a range of polaritons spanning a finite k window can be excited. This particularly applies to the PhP$_{[100]}$ polaritons on the thinnest flake with $d$ = 53 nm, wherein no reliable polariton dispersion data could be extracted as only a single fringe was observed. The effect of the finite bandwidth and temporal pulse width of the FEL pulses on the polariton interferometry analysis is discussed further in Supplementary Note S7. Quantifying figures of merit (FOM) of polaritons is crucial for their application in future technologies, as for example, in polaritonic resonators.[8,30] It is important to determine their associated decoherence lifetimes and performance-defining quality factors. To that end, we examined in detail the polariton propagation for all combinations of thicknesses and illumination frequencies used in our experiments. Representative profiles and their corresponding near-field images for PhP$_{[100]}$ ($\nu$ = 11.4 THz, $d$ = 131 nm) and PhP$_{[001]}$ ($\nu$ = 9.22 THz, $d$ = 53 nm) are shown in Figure 3d. The real-space profiles are corrected for the circular geometrical wave spreading by a factor[12,31] of $x^{-1/2}$, then are fitted with a damped sinusoidal function to obtain directly the modes' wavevectors Re($k$) and decay lengths $L$ = Im$(k)^{-1}$ which, combined with the previously determined $v_g$, directly yields the polariton lifetimes as $\tau = L/v_g$ (see Supplementary Note S2 for calculation details and $\tau$ values for all data points in Figure 3a). We estimate lifetimes of $\tau_{[001]}$ = 3.1 ± 0.4 ps (with associated propagation quality factor of Q$_{[001]}$ = Re(k)/Im(k) = 7.4 ) and $\tau_{[100]}$ = 9 ± 4 ps (with Q$_{[100]}$ = 4.3), for PhP$_{[001]}$ and PhP$_{[001]}$, respectively. Interestingly, the FOM for these THz polaritons show remarkable resemblance to the α-MoO$_3$ mid-IR polariton counterparts.[12,18,21] That is, they exhibit a very low-loss character, as evidenced by their exceptionally long lifetimes, while possessing low group velocities.

In summary, we provide a robust platform for control and confinement of long-wavelength THz radiation in nanoscale dimensions by exploiting phonon polaritonic excitations in the hyperbolic van der Waals crystal α-MoO$_3$. We thereby significantly extend the known range of ultra-low-loss polaritonic bands, which, moreover, feature in-plane hyperbolic propagation.

Our findings should open new avenues in the field of vdW heterostructuring, such as to enhance light absorption in photodetectors[32] or surface-enhanced spectroscopies,[33] increasing the efficiency in THz frequency converters,[19] or enabling non-linear control of matter with moderate THz field transients,[34] without necessarily relying on conventionally employed metal antennas. In addition, by exploiting the in-plane hyperbolic dispersion, α-MoO₃ could serve as a promising building block for enhancing the photon extraction rates in quantum emitters.[35]

**Methods Section**

*Scattering-type scanning near-field optical microscopy (s-SNOM)*: Nanoscale imaging was performed with a home-built s-SNOM end-station integrated at the free-electron laser at the Helmholtz-Zentrum Dresden-Rossendorf, Germany. A metallized tip is oscillated at its resonance frequency ($\Omega \sim 160$ kHz) in the vicinity of the sample surface while being excited by the FEL radiation. The tip acts as an antenna, concentrating the electric fields at its apex, which interacts with the sample volume, thus modifying the tip-scattered near-field signal ($S$). The near-field signal scattering has a non-linear dependence on the tip-sample distance, thus generating detected signals composed of multiple harmonics of the tip frequency ($n\Omega$, wherein $n = 1,2,3…$). As the far-field background signal is linearly modulated, demodulation of the scattered signals is done at a higher harmonic of the tip oscillation frequency ($n \geq 2$), effectively suppressing the background far-field contribution.[36] Throughout this work, the second-harmonic ($n = 2$) near-field signal ($S_{2\Omega}$) demodulated with lock-in amplifier is used. The scattered signal is recorded using a self-homodyne technique, as described elsewhere.[36,37] For the polariton interferometry measurements of the RB₁, we used a liquid helium-cooled gallium-doped germanium photoconductive detector (QMC Instruments Ltd), whereas for the RB₃, we used a liquid nitrogen-cooled mercury-cadmium-telluride photoconductive detector (Judson Technologies LLC, Model J15D26 equipped with a thallium bromo-iodide window).

*Free-electron laser tuning and diagnostic*: While tabletop lasers excel as sources for s-SNOM measurements in the near-IR to the mid-IR, suitable sources for s-SNOM in the THz spectrum are not as easily achieved. Alternatively, radiation emitted from relativistic electrons can be extremely bright THz sources, either incoherent[38] or coherent.[39,40] Synchrotrons generate broadband radiation extending down to IR and THz frequencies, where they have been applied for s-SNOM.[41,42] Here, restrictions still exist in the detection of the weak tip-scattered signals in s-SNOM, which has so far limited broadband nanospectroscopy with synchrotrons to frequencies >9.6 THz.[41] Free Electron Lasers (FEL) offer the advantage of broad continuous tunability combined with an extremely narrow spectral bandwidth (~ 0.5 – 2.5 % $_{FWHM}$). The higher spectral brightness of the FEL compensates for the reduced detection sensitivity at THz frequencies.[43,44]

At the ELBE Center for High Power Radiation Sources, two FELs, collectively referred to as FELBE, operate over a spectral range of 1.2 – 60 THz (5 – 250 μm) with a repetition rate of 13 MHz. For this study, the U100 FEL was utilized and provided the necessary brightness and spectral range to image the THz polaritons in α-MoO$_3$. Due to the highly dispersive and relative narrow-bandwidth nature of the THz polaritons, we tuned the FEL to its lowest achievable bandwidth by decreasing the length of its optical resonator cavity below the condition of perfect synchronization with the electron bunches from the accelerator. This causes the optical pulse in the FEL resonator to lead the electron bunch slightly on each pass through the undulator, thus reducing the overlap of each electron bunch with the stored optical FEL pulse. The reduced interaction between the electron bunch and the optical pulse decreases the FEL power, and also leads to lengthening of the FEL pulse, and a commensurate narrowing of the spectral bandwidth of the transform-limited FEL pulses.[45,46] Pulse bandwidths were kept at 0.51 - 0.97%$_{FWHM}$ throughout the experiments, as extracted from Figure 1b (main text). The latter pulse spectral diagnostic was performed with a calibrated grating spectrometer (SpectraPro 300, Acton Research Corp.).

*Full-wave numerical simulations and analytical isofrequency curves calculation*: The propagation of polaritons is fully determined by their isofrequency curve (IFC, a slice of the dispersion surface taken at a constant frequency). Therefore, to investigate how the PhPs propagate in α-MoO$_3$ in the THz frequency range, we calculated the analytical IFC by varying the angle α in supplementary Equation (S8) from 0 to $2\pi$ for a fixed incident frequency (black continuous lines in Figure 1d of the main text). To better visualize such propagation properties, we also performed full-wave electromagnetic simulations (COMSOL Multiphysics) to obtain the vertical component of the electric field spatial distribution, Re($E_z(x,y)$) (color plot in Fig 1d of the main text). To do this, α-MoO$_3$ structures were modelled as biaxial slabs[24,47] on top of high-resistivity float-zone Si substrates, in which PhPs were launched by vertically-oriented point electric dipole sources placed on top of the structure.

**Supporting Information**
Supporting Information is available from the Wiley Online Library or from the author.

**Acknowledgements**
The authors are grateful to Sergey Kovalev (HZDR, Dresden) for providing a few optical components, and the substrates used in this work. We also thank Sven Reitzig (TU Dresden) for the initial µ-Raman spectroscopy characterization, and the ELBE accelerator scientists and operators who assisted all FEL beamtimes #19201678. In addition, we acknowledge the efforts of Sergio Palacio Vega on extracting the THz dielectric function of α-MoO$_3$. T.V.A.G.O., T.N., L.W., M.O., S.C.K., and L.M.E. acknowledge the financial support by the German Federal Ministry of Education and Research (BMBF, Project Grant N$^{os}$ 05K16ODA, 05K16ODC, 05K19ODA and 05K19ODB) and by the Deutsche Forschungsgemeinschaft (DFG, German Research Foundation) under Germanýs Excellence Strategy through Würzburg-Dresden Cluster of Excellence on Complexity and Topology in Quantum Matter - ct.qmat (EXC 2147, project-id 390858490). G.Á-P. and J.T.-G. acknowledge the support through the Severo Ochoa Program from the Goverment of the Principality of Asturias (grants PA-20-PF-BP19-053 and PA-18-PF-BP17-126, respectively). P.A-G. acknowledges support from the European Research Council under Starting Grant 715496, 2DNANOPTICA. A.Y.N. acknowledges the Spanish Ministry of Science, Innovation and Universities (national project N° MAT201788358-C3-3-R). E. J. H. Lee acknowledges financial support from the Spanish Ministry of Science and Innovation, through the "María de Maeztu" Programme for Units of Excellence in R&D (CEX2018-000805-M) and the Ramón y Cajal grant RYC-2015-17973.

**References**

[1] T. Feurer, J. C. Vaughan, K. A. Nelson, *Science (80-. ).* **2003**, *299*, 374.

[2] A. J. Huber, B. Deutsch, L. Novotny, R. Hillenbrand, *Appl. Phys. Lett.* **2008**, *92*, 2.

[3] J. D. Caldwell, L. Lindsay, V. Giannini, I. Vurgaftman, T. L. Reinecke, S. A. Maier, O. J. Glembocki, *Nanophotonics* **2015**, *4*, 44.

[4] J. B. Khurgin, *Nat. Nanotechnol.* **2015**, *10*, 2.

[5] D. N. Basov, M. M. Fogler, F. J. Garcia de Abajo, *Science (80-. ).* **2016**, *354*, aag1992.

[6] J. Taboada-Gutiérrez, G. Álvarez-Pérez, J. Duan, W. Ma, K. Crowley, I. Prieto, A. Bylinkin, M. Autore, H. Volkova, K. Kimura, T. Kimura, M.-H. Berger, S. Li, Q. Bao, X. P. A. Gao, I. Errea, A. Y. Nikitin, R. Hillenbrand, J. Martín-Sánchez, P. Alonso-González, *Nat. Mater.* **2020**, 0.

[7] Y. Wu, Q. Ou, Y. Yin, Y. Li, W. Ma, W. Yu, G. Liu, X. Cui, X. Bao, J. Duan, G. Álvarez-Pérez, Z. Dai, B. Shabbir, N. Medhekar, X. Li, C.-M. Li, P. Alonso-González, Q. Bao, *Nat. Commun.* **2020**, *11*, 2646.

[8] J. D. Caldwell, A. V. Kretinin, Y. Chen, V. Giannini, M. M. Fogler, Y. Francescato, C. T. Ellis, J. G. Tischler, C. R. Woods, A. J. Giles, M. Hong, K. Watanabe, T. Taniguchi, S. A. Maier, K. S. Novoselov, *Nat. Commun.* **2014**, *5*, 1.

[9] E. Yoxall, M. Schnell, A. Y. Nikitin, O. Txoperena, A. Woessner, M. B. Lundeberg, F. Casanova, L. E. Hueso, F. H. L. Koppens, R. Hillenbrand, *Nat. Photonics* **2015**, *9*, 674.

[10] X. Lin, Y. Yang, N. Rivera, J. J. López, Y. Shen, I. Kaminer, H. Chen, B. Zhang, J. D. Joannopoulos, M. Soljačić, *Proc. Natl. Acad. Sci.* **2017**, *114*, 201701830.

[11] X. Lin, B. Zhang, *Laser Photon. Rev.* **2019**, *13*, 1900081.

[12] W. Ma, P. Alonso-González, S. Li, A. Y. Nikitin, J. Yuan, J. Martín-sánchez, S. Sriram, K. Kalantar-zadeh, S. Lee, R. Hillenbrand, Q. Bao, J. Taboada-gutiérrez, I. Amenabar, P. Li, S. Vélez, C. Tollan, Z. Dai, Y. Zhang, S. Sriram, K. Kalantar-zadeh, S. Lee, R. Hillenbrand, Q. Bao, *Nature* **2018**, *562*, 557.

[13] J. Duan, N. Capote-Robayna, J. Taboada-Gutiérrez, G. Álvarez-Pérez, I. Prieto, J. Martín-Sánchez, A. Y. Nikitin, P. Alonso-González, *Nano Lett.* **2020**, *20*, 5323.

[14] G. Hu, Q. Ou, G. Si, Y. Wu, J. Wu, Z. Dai, A. Krasnok, Y. Mazor, Q. Zhang, Q. Bao, C.-W. Qiu, A. Alù, *Nature* **2020**, *582*, 209.

[15] M. Chen, X. Lin, T. H. Dinh, Z. Zheng, J. Shen, Q. Ma, H. Chen, P. Jarillo-Herrero, S. Dai, *Nat. Mater.* **2020**, DOI 10.1038/s41563-020-0732-6.

[16] Z. Zheng, F. Sun, W. Huang, J. Jiang, R. Zhan, Y. Ke, H. Chen, S. Deng, *Nano Lett.* **2020**, *20*, 5301.

[17] S. Dai, Z. Fei, Q. Ma, A. S. Rodin, M. Wagner, A. S. McLeod, M. K. Liu, W. Gannett, W. Regan, K. Watanabe, T. Taniguchi, M. Thiemens, G. Dominguez, A. H. Castro Neto, A. Zettl, F. Keilmann, P. Jarillo-Herrero, M. M. Fogler, D. N. Basov, *Science (80-. ).* **2014**, *343*, 1125.

[18] Z. Zheng, J. Chen, Y. Wang, X. Wang, X. Chen, P. Liu, J. Xu, W. Xie, H. Chen, S. Deng, N. Xu, *Adv. Mater.* **2018**, *30*, 1.

[19] H. A. Hafez, S. Kovalev, J. C. Deinert, Z. Mics, B. Green, N. Awari, M. Chen, S. Germanskiy, U. Lehnert, J. Teichert, Z. Wang, K. J. Tielrooij, Z. Liu, Z. Chen, A. Narita, K. Müllen, M. Bonn, M. Gensch, D. Turchinovich, *Nature* **2018**, *561*, 507.

[20] S. Schlauderer, C. Lange, S. Baierl, T. Ebnet, C. P. Schmid, D. C. Valovcin, A. K. Zvezdin, A. V. Kimel, R. V. Mikhaylovskiy, R. Huber, *Nature* **2019**, *569*, 383.

[21] Z. Zheng, N. Xu, S. L. Oscurato, M. Tamagnone, F. Sun, Y. Jiang, Y. Ke, J. Chen, W. Huang, W. L. Wilson, A. Ambrosio, S. Deng, H. Chen, *Sci. Adv.* **2019**, *5*, eaav8690.

[22] W. Dong, R. Qi, T. Liu, Y. Li, N. Li, Z. Hua, Z. Gao, S. Zhang, K. Liu, J. Guo, P. Gao, **2019**, 1.

[23] L. Seguin, M. Figlarz, R. Cavagnat, J. C. Lassègues, *Spectrochim. Acta Part A Mol. Biomol. Spectrosc.* **1995**, *51*, 1323.

[24] G. Álvarez-Pérez, T. G. Folland, I. Errea, J. Taboada-Gutiérrez, J. Duan, J. Martín-

Sánchez, A. I. F. Tresguerres-Mata, J. R. Matson, A. Bylinkin, M. He, W. Ma, Q. Bao, J. I. Martín, J. D. Caldwell, A. Y. Nikitin, P. Alonso-González, *Adv. Mater.* **2020**, *32*, 1908176.

[25] J. Duan, R. Chen, J. Li, K. Jin, Z. Sun, J. Chen, *Adv. Mater.* **2017**, *29*, 1.

[26] A. Fali, S. T. White, T. G. Folland, M. He, N. A. Aghamiri, S. Liu, J. H. Edgar, J. D. Caldwell, R. F. Haglund, Y. Abate, *Nano Lett.* **2019**, *19*, 7725.

[27] S. Dai, J. Quan, G. Hu, C. W. Qiu, T. H. Tao, X. Li, A. Alù, *Nano Lett.* **2019**, *19*, 1009.

[28] J. Chen, M. Badioli, P. Alonso-González, S. Thongrattanasiri, F. Huth, J. Osmond, M. Spasenović, A. Centeno, A. Pesquera, P. Godignon, A. Zurutuza Elorza, N. Camara, F. J. G. de Abajo, R. Hillenbrand, F. H. L. Koppens, *Nature* **2012**, *487*, 77.

[29] A. J. Sternbach, S. Latini, S. Chae, H. Hübener, U. De Giovannini, Y. Shao, L. Xiong, Z. Sun, N. Shi, P. Kissin, G. X. Ni, D. Rhodes, B. Kim, N. Yu, A. J. Millis, M. M. Fogler, P. J. Schuck, M. Lipson, X. Y. Zhu, J. Hone, R. D. Averitt, A. Rubio, D. N. Basov, *Nat. Commun.* **2020**, *11*, 1.

[30] F. J. Alfaro-Mozaz, P. Alonso-González, S. Vélez, I. Dolado, M. Autore, S. Mastel, F. Casanova, L. E. Hueso, P. Li, A. Y. Nikitin, R. Hillenbrand, *Nat. Commun.* **2017**, *8*, 15624.

[31] A. Woessner, M. B. Lundeberg, Y. Gao, A. Principi, P. Alonso-González, M. Carrega, K. Watanabe, T. Taniguchi, G. Vignale, M. Polini, J. Hone, R. Hillenbrand, F. H. L. Koppens, *Nat. Mater.* **2015**, *14*, 421.

[32] S. Castilla, B. Terrés, M. Autore, L. Viti, J. Li, A. Y. Nikitin, I. Vangelidis, K. Watanabe, T. Taniguchi, E. Lidorikis, M. S. Vitiello, R. Hillenbrand, K. J. Tielrooij, F. H. L. Koppens, *Nano Lett.* **2019**, *19*, 2765.

[33] M. Autore, P. Li, I. Dolado, F. J. Alfaro-Mozaz, R. Esteban, A. Atxabal, F. Casanova, L. E. Hueso, P. Alonso-González, J. Aizpurua, A. Y. Nikitin, S. Vélez, R. Hillenbrand, *Light Sci. Appl.* **2018**, *7*, 17172.

[34] T. Kampfrath, K. Tanaka, K. A. Nelson, *Nat. Photonics* **2013**, *7*, 680.

[35] L. Shen, X. Lin, M. Y. Shalaginov, T. Low, X. Zhang, B. Zhang, H. Chen, *Appl. Phys. Rev.* **2020**, *7*, 021403.

[36] B. Knoll, F. Keilmann, *Opt. Commun.* **2000**, *182*, 321.

[37] L. Wehmeier, D. Lang, Y. Liu, X. Zhang, S. Winnerl, L. M. Eng, S. C. Kehr, *Phys. Rev. B* **2019**, *100*, 035444.

[38] W. D. Duncan, G. P. Williams, *Appl. Opt.* **1983**, *22*, 2914.

[39] W. M. Dennis, *J. Opt. Soc. Am. B* **1989**, *6*, 1045.

[40] G. L. Carr, M. C. Martin, W. R. McKinney, K. Jordan, G. R. Neil, G. P. Williams, *Nature* **2002**, *420*, 153.

[41] O. Khatib, H. A. Bechtel, M. C. Martin, M. B. Raschke, G. L. Carr, *ACS Photonics* **2018**, *5*, 2773.

[42] I. D. Barcelos, H. A. Bechtel, C. J. S. de Matos, D. A. Bahamon, B. Kaestner, F. C. B. Maia, R. O. Freitas, *Adv. Opt. Mater.* **2020**, *8*, 1.

[43] F. Kuschewski, H.-G. von Ribbeck, J. Döring, S. Winnerl, L. M. Eng, S. C. Kehr, *Appl. Phys. Lett.* **2016**, *108*, 113102.

[44] L. Wehmeier, T. Nörenberg, T. V. A. G. de Oliveira, J. M. Klopf, S.-Y. Yang, L. W. Martin, R. Ramesh, L. M. Eng, S. C. Kehr, *Appl. Phys. Lett.* **2020**, *116*, 071103.

[45] A. M. MacLeod, X. Yan, W. A. Gillespie, G. M. H. Knippels, D. Oepts, A. F. G. van der Meer, C. W. Rella, T. I. Smith, H. A. Schwettman, *Phys. Rev. E* **2000**, *62*, 4216.

[46] S. Regensburger, S. Winnerl, J. M. Klopf, H. Lu, A. C. Gossard, L. Fellow, S. Preu, *IEEE Trans. Terahertz Sci. Technol.* **2019**, *9*, 262.

[47] G. Álvarez-Pérez, K. V. Voronin, V. S. Volkov, P. Alonso-González, A. Y. Nikitin, *Phys. Rev. B* **2019**, *100*, 235408.